\documentclass[aps,prl,floats,twocolumn,showpacs,superscriptaddress]{revtex4}

\usepackage{graphics}
\usepackage{amsmath}
\usepackage{amsfonts}
\usepackage{amssymb}
\usepackage{times}
\usepackage{graphicx}% Include figure files
\usepackage{dcolumn}% Align table columns on decimal point
\usepackage{bm}% bold math
\begin{document}

\preprint{APS/123-QED}

\title{Diffusion Dynamics and Optimal Coupling in Directed Multiplex Networks}% Force line breaks with \\

\author{Alejandro Tejedor}
  \email{alej.tejedor@gmail.com}
\author{Anthony Longjas}%
\author{Efi Foufoula-Georgiou}%
\affiliation{%
 Department of Civil and Environmental Engineering, University of California, Irvine, Irvine, CA 92697, USA\\
}%

\author{Tryphon Georgiou}
\affiliation{
 Department of Mechanical and Aerospace Engineering, University of California, Irvine, Irvine, CA 92697, USA\\
}%

\author{Yamir Moreno}
\affiliation{%
 Institute for Biocomputation and Physics of Complex Systems (BIFI), Universidad de Zaragoza, 50018 Zaragoza, Spain\\
}%
\affiliation{
 Departamento de F\'isica Te\'orica, Universidad de Zaragoza, 50009 Zaragoza, Spain
}%
\affiliation{Institute for Scientific Interchange, ISI Foundation, Turin, Italy
}%

\date{\today}% It is always \today, today,

\begin{abstract}
We study the dynamics of diffusion processes acting on directed multiplex networks, i.e., coupled multilayer networks where at least one layer consists of a directed graph.  We reveal that directed multiplex networks may exhibit a faster diffusion at an intermediate degree of coupling than when the two layers are fully coupled. We use three simple multiplex examples and a real-world topology to illustrate the characteristics of the directed dynamics that give rise to a regime in which an optimal coupling exists. Given the ubiquity of both directed and multilayer networks in nature, our results could have important implications for the dynamics of multilevel complex systems towards optimality.
\end{abstract}

\maketitle

Multiplex networks are coupled multilayer networks where each layer consists of the same set of nodes but possibly different topologies and layers interact with each other only via counterpart nodes in each layer \citep{Domenico2013,Boccaletti2014, Kivela2014}. Multiplex networks have been shown useful for the study of diverse processes including social networks \citep{Cozzo2013,Li2015,Arruda2017}, transportation networks \citep{Aleta2017}, and biochemical networks \citep{Cozzo2012,Battiston2017}, among others.  Recent studies have shown that the coupling of the layers in a multiplex network can result in emergent structural \cite{Cozzo2016} and dynamical behavior such as enhanced diffusion (superdiffusion) \citep{Gomez2013}, increased resilience to random failure \citep{Domenico2014}, and emergence of critical points in the dynamics of coupled spreading processes \citep{Cozzo2013,Arruda2017}. These richer dynamics arise as a direct consequence of the emergence of more paths between every pair of nodes brought about by layer switching via an interlayer link.

Most of the theory for multiplex networks has been developed when all layers consist of undirected networks \citep{Domenico2013}.  However, more often than not, real social, biological and natural networks are structurally directed. Additionally, even if the underlying topology is not directed, the functional and dynamical connectivity of undirected networks is often directed due to gradients or the directionality in the flow of mass or information.  These processes include geophysical processes on tributary river networks \citep{Rodriguez-Iturbe1997,Zaliapin2010,Marra2014,Czuba2014} and  river delta channel networks \citep{Smart1971,Tejedor2015a,Tejedor2015b,Tejedor2016,Passalacqua2017,Tejedor2017}, food webs \citep{Milo2002,Krause2003,Pilosof2017}, gene regulation networks \citep{ShenOrr2002,Milo2002} and social dynamics \cite{Borge2011} to name a few.  In this paper, we study diffusion processes on directed multiplex networks, defined here as multiplex networks wherein the connectivity of at least one of the layers forms a directed graph.  We document a non-monotonic increase in the rate of convergence to the steady state as a function of the degree of coupling between layers.  We uncover that due to the directionality of (at least one of) the layers in a directed multiplex network an {\it optimal coupling regime} can emerge where transport processes are enhanced and diffusion is faster than when layers are fully coupled. 

Let $\mathbf{x}(t)$ represent the $N \times 1$ vector of concentration associated with the $N$ nodes of a network at time $t$ (throughout, vectors are thought of as column vectors). The diffusion dynamics on an undirected single layer network (monoplex) can be described by
\begin{equation}
\dot{\mathbf{x}}(t)=-D\cdot L{\mathbf{x}}(t),
\label{DiffEq}
\end{equation} 
\noindent where the vector $\dot{\mathbf{x}}(t)$ is the temporal derivative of $\mathbf{x}(t)$ and $D$ is a scalar that represents the diffusion constant.  The $L=[l_{ij}]^N_{i,j=1}$ matrix is the Laplacian of the network, which is defined as $L=S-W$. Here, the $N \times N$ matrix $W$ represents the weighted adjacency matrix, whose entries $w_{ij} = w_{ji} \in \mathbb{R}^+$ represent the strength of the connectivity between nodes $i$ and $j$, and $S$ is the $N \times N$ diagonal matrix with diagonal entries $s_{ii}=\sum^N_{j=1}w_{ij}$. For an unweighted network where the entries of $W$ are binary ($w_{ij} =1$ if there exists a link between nodes $i$ and $j$, and 0 otherwise), the matrix $S$ is called the degree matrix and its diagonal entries, $s_{ii}$ correspond to the number of links connected to node $i$, and the Laplacian is known as the combinatorial Laplacian (see \cite{Masuda2016} and references therein).

The negative Laplacian $-L$ (or more generally, $-D \cdot L$, when the diffusion constant $D \neq1 $) can be interpreted as the transition-rate matrix of a Continuous Time Markov Chain (CTMC); its entries $l_{ij}$ represent the rate at which the transition from node $i$ to node $j$ takes place \cite{Norris97,Masuda2016}. The dynamics of the corresponding Continuous Time Random Walk (CTRW), is governed by 
\begin{equation}
 \dot{\mathbf{p}}(t)^T=-{\mathbf{p}}(t)^TL,
\label{CTMCEq}
\end{equation} 
where $( )^T$ denotes transposition and  $\mathbf{p}(t)$ is a vector whose $i$-$th$ component represents the probability that the CTRW visits node $i$ at time $t$.  Note that for undirected networks, $L$ is symmetric, and therefore Eq. \ref{DiffEq} and Eq. \ref{CTMCEq} are equivalent for $D = 1$. Thus, the stationary distribution $(\dot{\mathbf{p}}(t)^T=0)$ of the CTRW is identical to the stationary solution of the diffusion process  ${\dot{\mathbf{p}}_s}^T=\frac{1}{N}(1,1,1,\cdots,1)^T$, which is the eigenvector of $L$ that corresponds to the zero eigenvalue. The characteristic time scale of convergence to that solution is inversely proportional to the smallest nonzero eigenvalue, $\lambda_2$, of $L$ assuming that the graph is connected $(\lambda_2>0)$ \citep{Syncreview08,Almendral2007}. 

For directed networks, the weighted adjacency matrix $W$ is not symmetric, i.e., the strength $w_{ij}$ of the directed link starting at node $i$ and ending at node $j$ may differ from $w_{ji}$.  Consequently, the Laplacian is not symmetric either.  Two different Laplacians can be naturally defined by extension of the undirected definition: (i) Out-Lapacian, $L^{out} = S^{out}-W$ and (ii) the In-Laplacian, $L^{in} = S^{in}-W$, where entries of the diagonal in- and out- strength matrices are defined as $s_{ii}^{out}=\sum_{j=1}^N w_{ij}$  and $s_{jj}^{in}=\sum_{i=1}^N w_{ij}$, respectively.  Both Laplacians (in- and out-) have $\mathbf{1}$ (column vector with all entries equal to 1) as the eigenvector corresponding to the zero eigenvalue, making it tempting to assume that  $\dot{\mathbf{x}}(t)=-D\cdot L^{out}{\mathbf{x}}(t)$ as the equation governing the diffusion process in directed networks.  However, due to the asymmetry of $L^{out}$, ${\bf 1}^T$ is not in general a left eigenvector of $L^{out}$ for the zero eigenvalue, and therefore continuity (conservation of mass) is not guaranteed in the process $(\frac{d}{dt}\sum x_i = \frac{d}{dt} \mathbf{1x} \neq 0 )$.  On the other hand, it is meaningful to interpret $-L^{out}$ as the transition rate matrix of a CTMC in a directed network, where the dynamics of the random walk is described by
\begin{equation}
 \dot{{\bf p}}(t)^T=-{\bf{p}}(t)^TL^{out}.
\label{CTMCOutEq}
\end{equation} 
In this case, continuity of the process (conservation of probabilities/mass) is assured since {\bf 1} is a right eigenvector corresponding to the zero eigenvalue of $-L^{out}$.  Furthermore, if the directed network is strongly connected (i.e., there exists a directed path between every pair of nodes) a unique stationary distribution of probability ${\bf p}_s^T$ (unique left eigenvector of the $-L^{out}$ corresponding to eigenvalue zero) is a probability vector (the Perron-Frobenius theorem guarantees that all the vector entries are positive given that the network is strongly connected),
\begin{equation}
 {\bf p}(t)^TL^{out}=0.
\label{CTMCOutSSEq}
\end{equation} 
The spectra of $L^{out}$ is in general complex, and the convergence towards the stationary distribution is exponential (asymptotic) with rate Re$(\lambda_2)$ where $\lambda_2$ is the eigenvalue with the smallest nonzero real part \citep{Lodato2007,Masuda2016}.

\citet{Gomez2013} generalized the study of the characteristic time scale of diffusive processes to (undirected) multilayer networks for different degrees of coupling across two different layers. The three main assumptions in \cite{Gomez2013} were: (1) the same set of nodes, albeit with different connectivity, forms the networks at each layer; (2) the connectivity in each layer (intralayer connectivity) consists of undirected networks forming a single-connected component; and (3) the interlayer connectivity consists of links between counterpart nodes in the different layers, i.e., multiplex.  The authors showed that the characteristic time of convergence of the diffusion process $\tau$ is inversely proportional to the smallest nonzero eigenvalue ${\Lambda_2}$ of the Laplacian of the multiplex, called supra-Laplacian $\mathcal{L}$ (i.e.  $\tau\sim \frac{1}{\Lambda_2}$). For the case of an undirected multiplex made up of two layers, with diffusion coefficients $D_1$ and $D_2$ for the respective layers, and an interlayer diffusion coefficient $D_x = D_{12} = D_{21}$, it can be shown that there are different asymptotic behaviors that depend on the relation between the diffusion coefficients $D_x$ and $D_1, D_2$. Interesting enough, there is a superdiffusion regime for some multiplex configurations, where for high values of coupling, diffusion can be faster in the multiplex than in any of the individual layers.

We next show that the heterogeneity introduced by the edge directionality of directed multiplex networks modifies substantially the behavior of diffusion processes on top of them. In this case, at an intermediate degree of coupling, multiplex may exhibit a faster diffusion than when the two layers are fully coupled (which by extrapolation of the undirected case \citep{Gomez2013} would have been expected to have the fastest rates). We refer to this region of coupling as the {\it optimal regime} and show that, within it, there is an {\it optimal coupling} for which the multiplex achieves the fastest diffusion. To this end, we study the characteristic time scale of convergence to the steady state of a CTRW acting on different directed multiplex networks consisting of (without loss of generality) two layers, wherein the intralayer connectivity is represented by unweighted graphs (i.e., we use combinatorial Laplacian). The supra-Laplacian is defined similarly to the undirected scenario,
\begin{equation}
\mathcal{L}^{out}= \begin{pmatrix}  D_1L_1^{out}+D_xI & -D_xI \\ -D_xI &  D_2L_2^{out}+D_xI \end{pmatrix}. 
\label{LaplacianMultiplexDirected}
\end{equation} 
Notice that $D_1$, $D_2$, and $D_x$ can be interpreted for a CTRW as scalars that control the relative speed of a walker in each layer and between layers, respectively.

\begin{figure*}[!t]
\centering
\includegraphics[width=0.9\textwidth, height=3.5in]{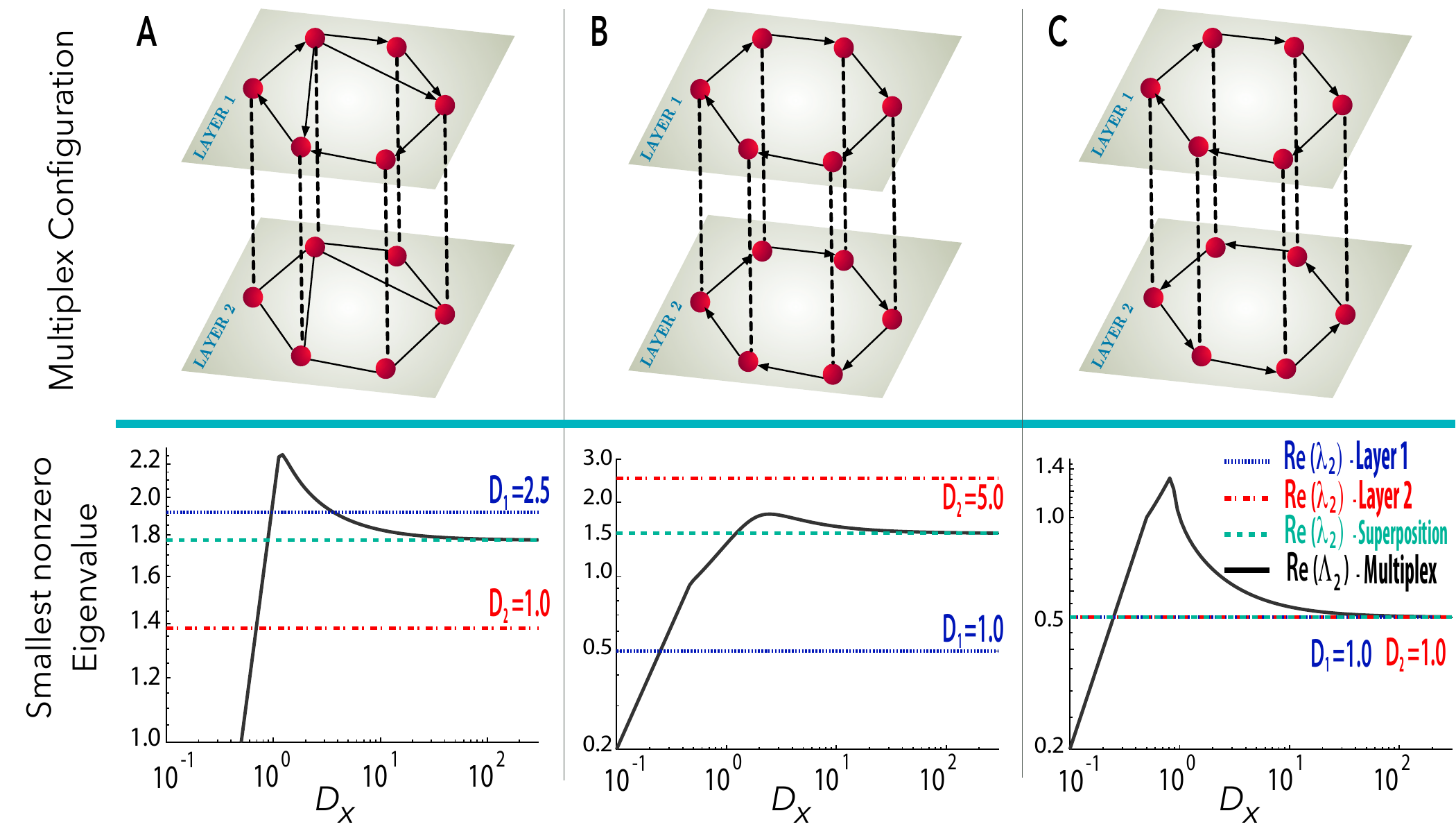}
\caption{(color online) The top panels depict the three synthetic multiplex networks with two coupled layers discussed in the text: (a) corresponds to Multiplex-1; (b) to the Multiplex-2 configuration; and (c) to the Multiplex-3 architecture. Bottom panels show the behavior of the smallest (in terms of its real part) nonzero eigenvalue Re$(\Lambda_2)$ of the supra-Laplacian $\mathcal{L}^{out}$ as a function of interlayer coupling, $D_x$.  See the text for further details.}
\label{Fig1}
\end{figure*}

The top panels in Fig.~\ref{Fig1} depict three multiplex networks, which are chosen to exemplify different types of heterogeneities that can appear when directed networks are considered: (a) {\it Multiplex-1} consists of a directed network (layer 1) and its undirected counterpart (layer 2) -- this example serves to illustrate the effect of directionality in only one of the layers; (b) {\it Multiplex-2} is composed of the same directed network in the two layers -- this example serves to show the effect of different rates ($D_1$ vs. $D_2$) at the different layers;  and (c) {\it Multiplex-3} where layer 1 contains the reverse directed network (opposite directionality of all the links) of the network present in layer 2 -- this example serves to illustrate the effect of directionality when the topology and rates of transition are the same. Note that for all three multiplex the interlayer links are undirected.

The bottom panels in figure~\ref{Fig1} present the smallest nonzero eigenvalue (in terms of its real part), Re$(\Lambda_2)$, of the  $\mathcal{L}^{out}$ as proxy of the time scale of convergence to the steady state $\tau$ (recall that $\tau\sim \frac{1}{Re(\Lambda_2)}$) when the coupling among layers (via $D_x$) is varied for each of the multiplex shown. The specific values of $D_1$ and $D_2$ used to reproduce the results are annotated in the figure.  Results from further exploration of these parameters are shown in the Supplemental Material.  Before discussing the specifics of each panel in Fig. \ref{Fig1}, the existence of four distinct regimes stand out: (1) {\it Linear}: layers are effectively decoupled and $D_x$ is the limiting factor.  In this case, Re$(\Lambda_2)$ increases linearly as $2D_x$; (2) {\it Sublinear}, wherein $D_x$ is larger than the smallest non-zero eigenvalue (in terms of the real part) of the slowest layer (Re$(\lambda_2^{\it slow})$, where the superscript {\it slow} refers to the slowest layer).  In this scenario, the slowest layer becomes the limiting factor (with respect to the rate of convergence of the multiplex to achieve the stationary distribution) and therefore an increase in $D_x$ translates into a sublinear rate of increase for Re$(\Lambda_2)$.  (3) {\it  Optimal}, corresponding to the range of $D_x$ for which Re$(\Lambda_2)$ exceeds the value of Re$(\Lambda_2)$ for $D_x \to \infty$.  Within the optimal coupling regime, we can define the optimal coupling as the value of $D_x$ for which Re$(\Lambda_2)$ achieves the absolute maximum.  The optimal coupling occurs for values of $D_x$ in the vicinity of the smallest nonzero eigenvalue (in terms of the real part) of the fastest layer (Re$(\lambda_2^{\it fast})$, where the superscript {\it fast} refers to the fastest layer).  In this case, the speed of the {\it transport} within the fastest layer competes with the transport across layers, achieving a configuration where both layers contribute significantly to the total transport but conserving a relative degree of independence in their internal dynamics (i.e., not fully synchronized or decoupled). (4) {\it Asymptotic} wherein  $D_x \gg$ Re$(\lambda_2^{slow})$,Re$(\lambda_2^{fast})$. The two layers are completely coupled and the counterpart nodes in the different layers are fully synchronized, behaving as a single node.  

%\begin{figure}[!t]
%\centering
%\includegraphics[width=0.99\columnwidth]{Fig2.pdf}
%\caption{}
%\label{Fig2}
%\end{figure}

Not all four regimes are present for every multiplex configuration.  In fact, the linear and the asymptotic (even it could be argued sublinear as well) regimes were observed and defined for undirected multiplex in \cite{Gomez2013}.  However, the {\it optimal coupling regime} is a characteristic that solely applies to multiplex networks that have a directed network in at least one layer. The difference in the dynamics of directed and undirected multiplex emerges from the fact that the directionality of the links in directed (networks) layers allows faster exploration of the other nodes in the same layer (e.g., when a random walker leaves through a link, there is no way back through the same route). Note that this asymmetry in the path directionality is, in general, not advantageous with respect to achieving stationary states for diffusion-like processes.  However, directed networks, when integrated in a multiplex, can be catalysts of diffusion-like processes when a right balance between their coupling with other layers (high enough to have access, through different layers, to shortcuts that overcome the asymmetry of the paths) and the degree of independent dynamics they preserve (take advantage of their faster exploratory capability within the layer) exists. Notice that a complete coupling of the layers in a multiplex can be interpreted as a monoplex resulting in the superposition of the connectivity of the different layers. This effect would result in the removal of the constraint on path asymmetries, and therefore potential suboptimal times of convergence to steady state. 

The emergence of the optimal coupling is phenomenologically discussed  for each of the multiplex illustrated in Fig. \ref{Fig1}. For {\it Multiplex-1}, an optimal regime can be observed in the vicinity of $D_x = 1.2$ resulting from the access to different topological paths when the directed network is coupled to its undirected counterpart. In the case of the {\it Multiplex-2}, the networks are identical at the different layers, and therefore no new topological paths are created when the layers are combined in the multiplex.  However, different values of the $D_1$ and $D_2$ can create distinct dynamic paths (i.e., same topology but different rates of transport) giving rise to different gradients among counterpart nodes in the separate layers, resulting in the emergence of an optimal coupling for $D_x \sim 2.4$.  For this multiplex, a transition from linear to sublinear regime is also apparent when values of $D_x$ exceed Re$(\lambda_2^{slow} )=0.5$.  

The case of {\it Multiplex-3} is particularly interesting. In this case, an optimal coupling is observed for $D_x \sim 0.8$ resulting from access to different topological paths when the directed network is coupled to its reverse. This configuration serves as a word of caution, since in many instances, undirected networks are assumed if transitions in both directions of a link ($i\rightarrow j$ and $j \rightarrow i$) are possible. However, if the process is such that it operates with certain degree of independence in each direction, the properties of the dynamics of such a system can substantially differ from those of a fully undirected network. In fact, we show how the multiplex formed by the two reverse networks can function faster than each of the layers and also faster than the equivalent undirected network ($\lambda_2 =1$). 

In \cite{Gomez2013}, it was shown that in fully coupled undirected multiplex, diffusion can be faster than in the fastest of its individual layers (superdiffusion), since at high coupling, the multiplex can be approximated by the superposition network, where the availability of paths between nodes is increased, enhancing the efficiency of diffusive processes.  For a multiplex consisting of at least one directed network, there exist two regimes of coupling wherein a superdiffusion-like behavior can be observed: (1) {\it Asymptotic} -- equivalent to the one described for the undirected multiplex; and (2) {\it Optimal} -- where for an intermediate degree of coupling, the speed of the processes is higher than when the two layers are fully coupled (e.g., the asymptotic scenario). In this superdiffusion regime, diffusion may even be faster than in {\it any} of the individual layers. Thus, directed multiplex networks can exhibit superdiffusion within the optimal regime, even in scenarios where the asymptotic regime is not superdiffusive (e.g., Fig.~\ref{Fig1} bottom panels (c)). 

\begin{figure}[!t]
\centering
\includegraphics[width=0.99\columnwidth]{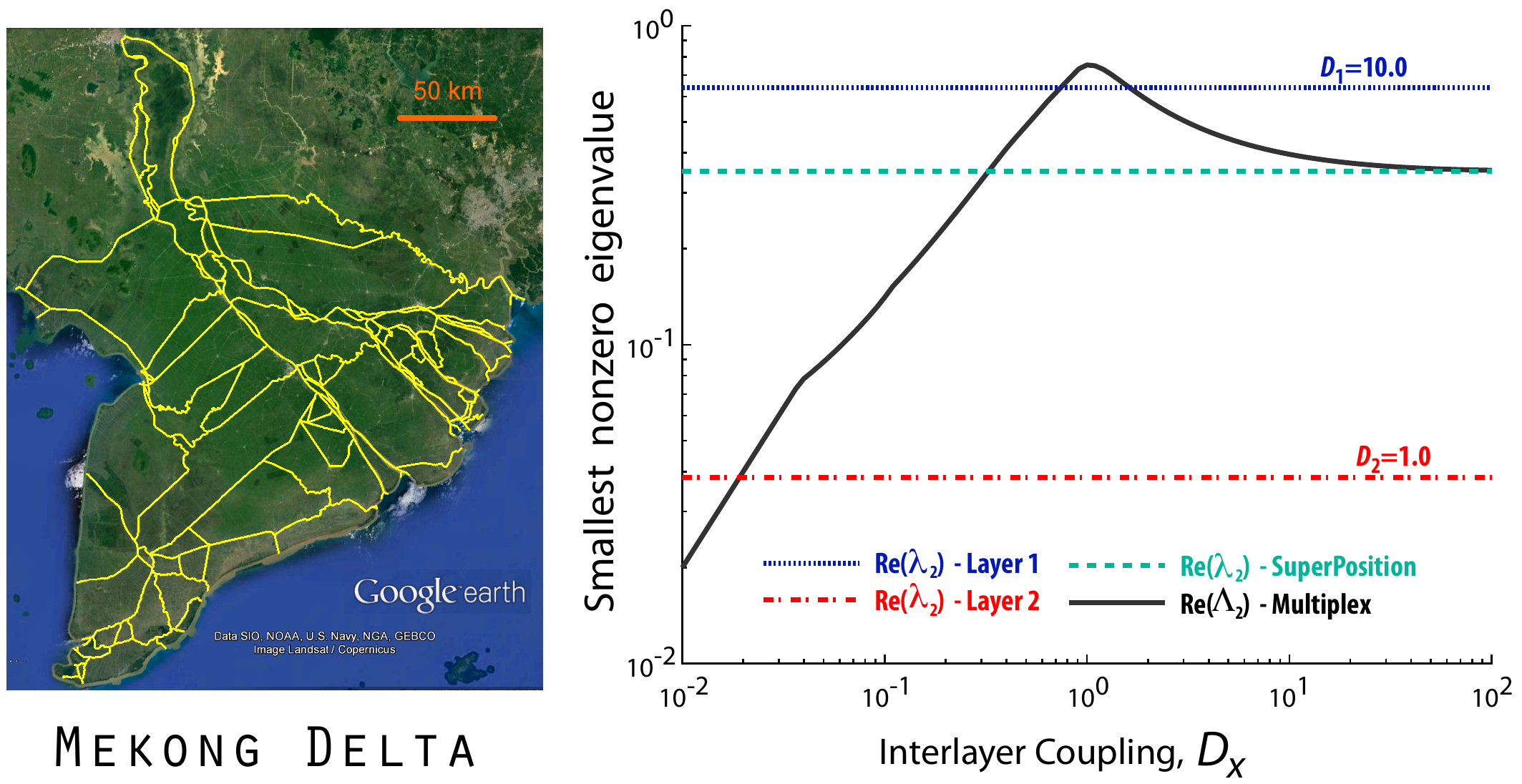}
\caption{(color online) The Mekong delta Multiplex \cite{note}. The left panel shows the Landsat image of the Mekong delta (Data Source: Google Earth (Data SIO, NOAA, U.S. Navy, NGA, GEBCO) Landsat/Copernicus imagery date: 12/13/2015), where the extracted channel network at resolution of 50 m is superimposed in yellow.  The right panel shows the behavior of the smallest (in terms of its real part) nonzero eigenvalue Re$(\Lambda_2)$ of the supra-Laplacian $\mathcal{L}^{out}$ corresponding to the Mekong delta multiplex as a function of $D_x$.  For intermediate coupling rates, $D_x \sim 1$, maximum Re$(\Lambda_2)$ is observed, i.e., the fastest rate of convergence to the steady-state solution (even faster than the asymptotic limit for fully coupled layers).}
\label{Fig2}
\end{figure}

Finally, we also illustrate the existence of the optimal coupling regime in a real-world network with a more complex topology.  Specifically, we use the channel network of the third largest delta on Earth, Mekong delta  (see Supplemental Material for network description and more information on the Mekong delta), to examine the characteristic time of convergence to steady state of a CTMC operating on a two layer multiplex:  Layer 1 consists of the directed channel network, where  the direction of the edges corresponds to the main direction of water flow;  and  Layer 2 consists of the undirected counterpart of the network in Layer 1. Note that the dynamics of this multiplex can be representative of transport processes that have both advective (operating on layer 1) and diffusive (operating on layer 2) components.   By varying the interlayer coupling $D_x$, we show in Fig.~\ref{Fig2} that an optimal coupling is observed for $D_x \sim 1$, which also exhibits superdiffusion.

In conclusion, we revealed the existence of an optimal coupling regime in which directed multiplex networks may exhibit a faster diffusion than in the asymptotic limit, when the different layers are fully coupled. We argue that it is precisely the directionality of paths that sets an anisotropic layout for the process, and combined with a balance of (1) significant connectivity across layers (making accessible paths in other layers) and (2) a degree of independence in the intralayer dynamics, can catalyze the overall system process.  Moreover, using three synthetic and one real multiplex networks, we illustrated different heterogeneities that can lead to a multiplex with an optimal coupling state and a new superdiffusion regime. Our results open-up new paths of research addressing questions such as whether natural complex systems self-organize to configurations where the optimal coupling is accessible to their dynamics.

\begin{acknowledgments} This work is part of the International BF-DELTAS project on ``Catalyzing action towards sustainability of deltaic systems" funded by the Belmont Forum (NSF grant EAR-1342944) and the
LIFE (Linked Institutions for Future Earth, NSF grant EAR-1242458). Partial support is also acknowledged from the Water Sustainability and Climate Program (NSF grant EAR1209402).  A.T. acknowledges financial support from the National Center for Earth-surface Dynamics 2 postdoctoral fellowship (NSF grant EAR-1246761). Y. M. acknowledges partial support from the Government of Arag\'on, Spain through a grant to the group FENOL, and by MINECO and FEDER funds (grant FIS2014-55867-P).
\end{acknowledgments}

%\bibliographystyle{apsrev4-1} % Tell bibtex which bibliography style to use
%\bibliography{Bib_MultiplexDirected}

\begin{thebibliography}{33}
\bibitem{note} Based on the channel network, we define a two layer multiplex, wherein Layer 1 consists of the directed channel network (the direction of the edges corresponds to the main direction of water flow) and Layer 2 consists of the undirected counterpart of the network in Layer 1.
\end{thebibliography}
\end{document}